\documentclass[preprint]{revtex4}
\begin{document}
%\tighten
\bibliographystyle{prsty} %Physical Review referencing style
\draft %makes pacs numbers print
\input epsf.sty %used to print eps files  

\title{Quantum Hydrodynamic Model for the enhanced moments
of Inertia of molecules in Helium Nanodroplets: Application to
SF$_6$}

\author{Kevin K. Lehmann}
\email{Lehmann@princeton.edu}
\author{Carlo Callegari}

\affiliation{Department of Chemistry, Princeton University, Princeton NJ 08544
USA}
\date{\today}

\begin{abstract} 
The increase in moment of inertia of SF$_6$ in helium nanodroplets is
calculated using the quantum hydrodynamic approach.  This required
an extension of the numerical solution to the hydrodynamic equation to
three explicit dimensions.  Based upon an expansion of the density in
terms of the lowest four Octahedral spherical harmonics, the
predicted increase in moment of inertia is $170 {\rm u \AA^2}$,
compared to an experimentally determined value of $310(10)\,{\rm u \AA^2}$,
{\it i.e.}, 55\% of the observed value.  The difference is likely in at least
part due to lack of convergence with respect to the angular expansion, but at
present we do not have access to the full densities from which a higher order
expansion can be determined.  The present results contradict those of 
Kwon {\it et al.}, J. Chem. Phys. {\bf 113}, 6469 (2000), who predicted
that the hydrodynamic theory predicted less than $10 \%$ of the observed
increase in moment of inertia.

\end{abstract}

\maketitle

There is considerable current interest in the spectroscopy of
atoms and molecules solvated in liquid Helium, particularly in
$^4$He nanodroplets~\cite{Toennies98}.  These provide
microscopic probes of the nature of this unique substance, the
only physical system whose equilibrium state remains a liquid
as $T \rightarrow 0$\,K.  Further, helium has properties that
make it an almost ideal `matrix' for the production and
characterization of novel chemical
species~\cite{Lehmann98}.  One such property is that even
highly anisotropic solutes give rotationally resolved spectra,
though with effective rotational moments of inertia several
times that of the isolated molecule in the gas
phase~\cite{Hartmann95,Callegari00b}.  Development of a
quantitative predictive theory for the enhanced effective
moments of inertia will be valuable for future use of the
spectroscopically observed rotational constants to help
determine the chemical carrier of an unassigned spectral
feature, as they are often used in gas phase spectroscopy.

It has been widely recognized that enhanced moments of inertia
arise from the kinetic energy of helium motion that is
correlated with the rotation of the solute.  In the case of
SF$_6$-He$_N$~\cite{Lee99}, a fixed frame, fixed node diffusion Monte
Carlo calculations on small clusters with $N = 8 - 20$ He atoms have
recovered rotational excitation spectra in excellent
agreement with that observed for SF$_6$ in He nanodroplet with
much larger number of helium atoms.  Such calculations do not
directly give any dynamical information, however, and thus
leave open the question of how to physically characterize the
helium motion.  Two quantitatively predictive dynamical models
have been put forth that invoke very
distinct types of helium motion~\cite{Kwon99b,Callegari99b}.  In
one~\cite{Kwon99b,Kwon00}, a proposed `nonsuperfluid fluid' density of helium
was calculated using Path Integral Monte Carlo (PIMC) methods, and
this density is assumed to rotate rigidly with
the molecule provided that the molecule-He potential is sufficiently
anisotropic relative to the induced rotational energy.  
To date, this two fluid model has been applied only to
SF$_6$ and OCS~\cite{Kwon99b,Kwon00}.  
For SF$_6$, the `two fluid model' gave a result
in excellent agreement with experiment~\cite{Hartmann95}. 
The predicted increase in the effective moment of inertia, $\Delta I$,
was calculated as $327 {\rm u \AA^2}$ compared
to a valued of $310(10)\,{\rm u \AA^2}$ calculated from the
observed rotational constants.
The level of agreement, fact, is far better than should be
expected given the considerably uncertainty in the relevant He-SF$_6$
interaction potential~\cite{Pack84}.  The two parameters that
determine the anisotropy of the He-SF$_6$ potential well have
reported values of $-0.6 \pm 0.3$ and $0.14 \pm 0.14$ in
Ref.~\cite{Pack84}.  

The second approach, published by ourselves and
coworkers~\cite{Callegari99b}, is based upon a hydrodynamic
treatment for the helium flow, which is assumed to maintain a
constant solvation density in the frame rotating with the
molecule and be ideal (aviscous and irrotational).   The
hydrodynamic approach was applied to a number of linear
molecules, and was found to be in good agreement for heavier
molecules, including OCS.  A key assumption of this work is that of adiabatic
following by which we mean that the helium density in the
molecular frame of a rotating molecule is the same as for a
rigid molecule, fixed in the laboratory frame.  It has recently
been established experimentally~\cite{Conjusteau00} that for
lighter molecules, in particular HCN and DCN, that adiabatic
following breaks down and explains at least the sign of the
error of the hydrodynamic theory in these cases.

In order to apply the hydrodynamic model, one needs the three
dimensional helium density around the solute, which is known to
be highly structured.  The earlier hydrodynamics work used
helium density functional theory (DF)~\cite{Casas95} to
estimate this quantity.  For systems with cylindrical or higher
symmetry, DF is many orders of magnitude computationally less
expensive than Quantum Monte Carlo Methods but introduces
additional uncertainties beyond the ever present uncertainty in
the Helium-Solute interaction potential.  In this work, the
helium density around a static SF$_6$ molecule, previously
published by Barnett and Whaley~\cite{Barnett93}, is used to
calculate the hydrodynamic contribution to the moment of
inertia of this molecule.   Because of the different symmetry
(O$_{\rm h}$ versus C$_{\infty \rm v}$) of this solute compared
to those treated previously, some changes had to be made in the
computational procedure.  Most significant, one degree of
freedom could not be removed by separation of variables (as
it could be for a cylindrically symmetric density) and, as a
result, the hydrodynamic equation for the velocity potential had
to be solved numerically in three degrees of freedom.   The
computational procedures used in the present case are described
in the next section and the results presented in the last
section.

\section{Hydrodynamic Calculations}

The hydrodynamic calculations require an estimate of the
three dimensional helium density around the solute molecule,
SF$_6$ in this case.  This density will be totally symmetric
in the point group of the molecule, O$_{\rm h}$ in this case.
This density, $\rho(r,\Omega)$, can be expanded in terms of the
spherical tensor operators:
\begin{equation}
\rho(r,\Omega) = \rho_0(r) + a_4(r)\,T_4(\Omega) + a_6(r)\,
T_6(\Omega) + a_8(r)\,T_8(\Omega) + \ldots
\label{eq:rho_expansion}
\end{equation}
where $\Omega = (\theta,\varphi)$ and $T_L$ are linear
combinations of the spherical harmonics,
$Y_{LM}$'s, that transform as $A_{1\rm g}$ in the $O_{\rm h}$
point group. In particular:
\begin{eqnarray}
T_4 &=& \sqrt{\frac{7}{12}}Y_{40} + \sqrt{\frac{5}{24}}
\left( Y_{40} + Y_{4-4} \right) \\
T_6 &=& \sqrt{\frac{1}{8}}Y_{60} - \frac{\sqrt{7}}{4}
\left( Y_{60} + Y_{6-4} \right) \\
T_8 &=& \frac{\sqrt{33}}{8}Y_{80} + \frac{\sqrt{42}}{24}
\left( Y_{84} + Y_{8-4} \right) 
+ \frac{\sqrt{390}}{48} \left( Y_{88} + Y_{8-8} \right)
\end{eqnarray}
Fox and Ozier~\cite{Fox70} have presented a
general procedure for to calculate these and higher harmonics.

\begin{figure}
\centerline{\epsfbox{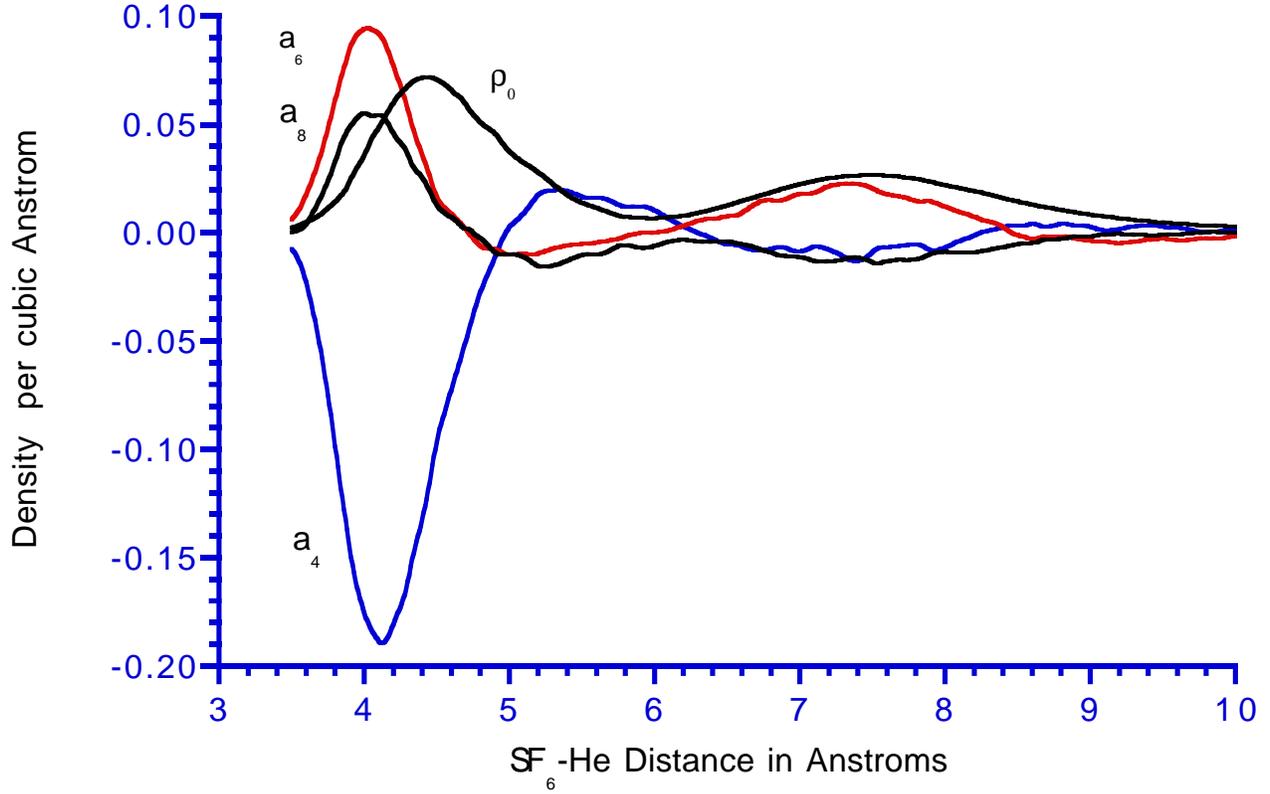}}
\vspace{1in}
\caption{The tensor densities,$\rho_0(r), a_4(r), a_6(r)$,
 and $a_8(r)$,  in ${\rm \AA}^{-3}$, as a function of the
He-SF$_6$ radial distance in $\rm \AA$}
\label{fig_a}
\end{figure}

Barnett and Whaley have calculated the helium density around
SF$_6$ using the diffusion Monte Carlo method.  Their paper
contains figures which give the radial isotropic density,
$\rho_0(r)$, and radial `cuts' of the density along the C$_2$,
C$_3$, and C$_4$ symmetry axes.  If we assume that the density
contains only the terms explicitly given in
Eq.~\ref{eq:rho_expansion}, then we can use the isotropic
density and the density cuts to determine the radial
coefficients
$a_4(r), a_6(r),$ and $a_8(r)$ by using the equations:
\begin{eqnarray}
\sqrt{\frac{21}{\pi}} a_4(r) &=& \frac{378}{143} \rho_0
-\frac{256}{143} \rho_2 -\frac{378}{143}\rho_3 
+ \frac{256}{143} \rho_4 \\
\sqrt{\frac{26}{\pi}} a_6(r) &=& -\frac{20}{55} \rho_0
-\frac{128}{55} \rho_2 +\frac{108}{55}\rho_3 
+ \frac{40}{55} \rho_4 \\
\sqrt{\frac{561}{\pi}} a_8(r) &=& -\frac{1680}{65} \rho_0
+\frac{768}{65} \rho_2 +\frac{432}{65}\rho_3 
+ \frac{480}{65} \rho_4 
\end{eqnarray}
In these equations, $\rho_i$ represents the radial density
along the C$_i$ symmetry axis.  The radial density functions
where determined by digitization of the images from the paper
by Barnett and Whaley for clusters of 69 Helium atoms, the
largest for which they reported the density cuts. 
These equations were derived by evaluation of the tensor
operators ($T_i$) along each of the symmetry axes and
then inverting the linear system that relates the calculated
density along each cut.
Figure~\ref{fig_a} shows the radial tensor densities
calculated from these density cuts.
It is apparent from this figure that the density in the
first solvation shell (which is the most important for
determination of the moment of inertia) is likely not
fully converged with the present truncation of the
density expansion.
Unfortunately, we were not able to obtain
the primary data from reference \cite{Barnett93}
which would have allowed the expansion to be carried to higher
order.

The assumption that the helium motion is irrotational 
($\mbox{\boldmath$\nabla$} \times {\bf v} = 0$) implies
that the velocity, 
${\bf v}$, can be written as the gradient of a scalar function,
${\bf v} = - \mbox{\boldmath$\nabla$}\phi$, where $\phi$ is
known as the velocity potential.
In order to calculate the increased moment of inertia
caused by motion of the helium, we will assume that the
molecule is undergoing classical rotation with angular
velocity $\omega$ around the $z$ axis.  The
equation of continuity gives the following equation:
\begin{equation}
\mbox{\boldmath$\nabla$} \cdot ( \rho \mbox{\boldmath$\nabla$}
\phi) = \frac{\partial \rho}{\partial t} = 
- (\mbox{\boldmath$\nabla$} \rho) \cdot \left( \omega 
\hat{\mbox{\boldmath$\varphi$}} \right)  \label{eq:hydro}
\end{equation}
where the second equality comes from the assumption that
the helium density remains constant in the frame rotating
with angular velocity $\omega$ around the $z$ axis.  The
boundary conditions on
$\phi$ are that the normal component of the fluid velocity
match the normal component of velocity of any surface.  
The solution to Eq.~\ref{eq:hydro} is linear in the
angular velocity and thus is solved numerically for unit
velocity.  Equating the helium kinetic energy with the
rotational kinetic energy, we can define $\Delta I_{\rm h}$ by:
\begin{eqnarray}
\frac{1}{2} \Delta I_{\rm h} \omega^2 &=&
\frac{1}{2} m_{\rm He} \int \rho \left| 
\mbox{\boldmath$\nabla$}\phi \right|^2 {\rm d}V  \label{ekin1}\\
 &=& \frac{1}{2} \, m_{\rm He} \left[
- \int \, \phi \, \left( \frac{\partial \rho}{\partial t} \right) dV
+ \int \, \rho \, \phi \, (\mbox{\boldmath$\nabla$} \phi) \cdot
d{\bf S} \right]
\label{ekin2}
\end{eqnarray}
The first equation holds for any velocity potential.  The
second has been derived from the first using vector 
identities and Eq.~\ref{eq:hydro}.  As such, these two
estimates need only be equal for $\phi$ that is a solution
to Eq.~\ref{eq:hydro}.
It can also be shown that the net orbital angular momentum
produced by the helium flow is, for the solution of
Eq.~\ref{eq:hydro}, ${\bf J}_{\rm He} = \Delta I_{\rm h}\,
\mbox{\boldmath$\omega$}$.

A numerical solution for $\phi$ was found on a grid of points in
a spherical coordinate system.  The three cartesian axes of
this coordinate system is aligned with three of the S-F bonds. 
The range of $r$ was selected between 3.5 and 10 $\rm \AA$ with
between 51 and 201 radial points.  The normal component of
$\phi$ was selected to be zero on both the inner and outer
constant $r$ surfaces.
$\phi$ must be invariant to reflection in the
$x,y$ plane, and thus we numerically restricted the solution to
the $\theta = \left[0,\pi/2\right]$ and
restrict the solution to have zero normal derivative on
the $x,y$ plane.  
Reflection in each of the four planes perpendicular to the $z$
axis is equivalent to change in the direction of rotation,
and thus leads to a change in the sign of $\phi$.  
As a result, the numerical solution for $\phi$ could
be restricted to the domain $\varphi = \left[0,\pi/4\right]$
and $\phi$ selected to be zero on each of the planes
$\varphi = 0 \, {\rm and} \, \pi/4$.  Between
41 and 161 equally spaced angular points were used in each of
the angular coordinates coordinates.  
Since $\phi$ or its normal derivative is equal to zero
on all the boundary surfaces, the surface integral
in Eq.~\ref{ekin2} is zero.  

The inhomogeneous partial differential Eq.~\ref{eq:hydro}
was converted to a finite difference equation in the grid of
points and solved by Gauss-Seidel iteration with successive
overrelaxation~\cite{Numerical_Recipes}.  These
equations (which involve $\ln \rho$ and its gradient)
become singular when $\rho = 0$.  As a result, the
density is bounded to remain above a threshold value,
which was selected as $\rho_{\rm min} = 10^{-5}\,
\rm \AA^{-3}$ in this work.  Solutions were iterated until the
mean squared change in
$\phi$ on the grid points was less than a fixed fraction
($10^{-10}$ in this work) of the mean squared value of $\phi$
on the grid points. For the largest grid used, this required
about 6000 iteration cycles with an overrelaxation parameter
of 1.4 ($\omega$ in the notation used in Numerical
Recipes~\cite{Numerical_Recipes}). The two integral estimates
for $\Delta I_{\rm h }$ are found to be 171 and 168 $\rm
u \AA^2$ respectively.  This can
be compared with a value of $\Delta I = 310(10)\,{\rm u \AA^2}$
inferred from the observed effective rotational constant of
SF$_6$ in $^4$He nanodroplets~\cite{Hartmann95}.

Kwon and Whaley~\cite{Kwon99b} reported the `superfluid'
densities along the same symmetry axes, as calculated using
PIMC and their proposed superfluid estimator.  These have
greatly decreased anisotropy compared to the total density. 
Repeating the hydrodynamic calculation using these densities
gave $\Delta I_{\rm h} = 12\,{\rm u \AA^2}$.
This can be compared to their `superfluid' contribution
of the moment of inertia increase of $22\,{\rm u \AA^2}$~\cite{Kwon00}.
These authors also claim to have calculated the hydrodynamic
contribution to the rotational constant used the total density
around SF$_6$ and report $\Delta I_{\rm h} = 31\,{\rm u \AA^2}$.
This value is 5.5 times smaller than the value we have calculated.
Based upon our extensive experience with similar hydrodynamic
calculations, the modest 50\% in hydrodynamic moment of
inertia is inconsistent with the considerably greater angular
anisotropy of the total density compared to their reported
`superfluid' density.

\section{Discussion}

The hydrodynamic prediction for the enhanced moment of inertia
of SF$_6$ in helium is $\sim 55\%$ of the experimentally
observed value.  This can be contrasted with the results
reported in \cite{Callegari99b}, where for heavier rotors, the
theory appeared to systematically overestimate the size of the
increased moment of inertia.  Similarly, an overestimate of the
hydrodynamic estimate for the increased moment of inertia was
found for a model problem of a planer rotor interacting with a
rigid ring of He atoms~\cite{Lehmann00b}.  One explanation for
the underestimate in the present case is that the density model
used has underestimated the true anisotropy of the helium
density, particularly in the first solvation shell where the
hydrodynamic kinetic energy density is highest.  This is
supported by the results presented in fig 1, where it is
demonstrated that the highest order anisotropy we have
retained is still of considerable size in this first shell. 
Since the $\varphi$ derivative of the density is the source
term in the hydrodynamic equation, addition of higher order
anisotropies is expected lead to increased hydrodynamic motion
and thus an increased estimate of $\Delta I_{\rm h}$.
This explanation for the limited success of the hydrodynamic
model in the present case can be tested by repeating the
hydrodynamic calculation using the full anisotropic helium
density calculated by Quantum Monte Carlo methods.

An alternative explanation, which cannot be ruled out at
present, is that some fraction of the helium motion is not 
'irrotational'.  The Kelvin minimum energy principle~\cite{Milne} states
that any `rotational' solution to the equation of continuity
will generate higher helium kinetic energy and thus He
contribution to the moment of inertia.  This would include the
hydrodynamics assumed by the `two fluid' model of Kwon and
Whaley~\cite{Kwon99b}. 

The fixed node, Diffusion Monte Carlo calculations of Lee {\it
et al.}\cite{Lee99} were in excellent agreement with the
experimental rotational excitation energies.  This suggest
that, barring cancelation of errors, that the nodal structure of
the wavefunction assumed in that work (which had the nodal
properties of the rigid rotor wavefunction for the SF$_6$)
should be a reasonable description of the true many body
wavefunction for this system. It is useful, therefore, to
examine how consistent the hydrodynamic model is with the
functional form assumed in that work.

In a future publication,
we will present a quantum derivation of the hydrodynamic
approach.  In that work, it will be shown
that Eq.~\ref{eq:hydro} arises from a variational optimization
of a one particle phase function that multiples the ground
state wavefunction that describes the helium in the frame that
rotates with the molecule.  The wavefunction for the
orientation of this axis system in space will be the rigid
rotor function for the molecule.  Thus, the presence of the
one particle phase functions will modify the nodal surfaces from
that assumed in the Lee {\it et al.} calculation.  The size of
the one particle phase argument will be $m_{\rm He} \phi /
\hbar$.  The maximum value for $\phi$ for our solution is $\sim
5 \AA^2 \omega$, and for the $J=1$ level we can approximate
$\omega = 4\pi B_{\rm eff}$, where $B_{\rm eff}$ is the
effective moment of inertia of SF$_6$ in helium, $1.04$\,GHz.
This gives a maximum hydrodynamic phase of $\approx 0.04$,
and thus it appears that this will lead of small changes in the
nodal properties.  We would like to point out that for the model
problem of the planer rotor coupled to a ring of Helium (which
can of course be solved exactly), the errors in the equivalent
fixed node approximation was of this same size and yet the DMC
estimate of the rotational excitation energy had errors of at
most a few percent.

\section{Acknowledgement}

The authors wish to acknowledge Prof. Giacinto Scoles for
many helpful discussions.  This work was supported by the
National Science Foundation and the Air Force Office of
Scientific Research.

%\bibliography{Helium}

\end{document}